\documentclass{elsart}
\usepackage{amsmath}
\usepackage{rotating}

\input{tcilatex}
\begin{document}

\begin{frontmatter}

\title{Asymmetric Conditional Volatility in International Stock Markets}

\author{Nuno B. Ferreira*, Rui Menezes* and Diana A. Mendes*}

\address{*Department of Quantitative Methods, IBS - ISCTE Business School, ISCTE,  Av. Forcas Armadas, 1649-025 Lisboa, Portugal, e-mail: nuno.ferreira@iscte.pt}

\begin{abstract}
Recent studies show that a negative shock in stock prices will generate more volatility than a positive shock of similar magnitude. The aim of this paper is to appraise the hypothesis under which the conditional mean and the conditional variance of stock returns are asymmetric functions of past information. We compare the results for the Portuguese Stock Market Index PSI 20 with six other Stock Market Indices, namely the SP 500, FTSE 100, DAX 30, CAC 40, ASE 20, and IBEX 35. In order to assess asymmetric volatility we use autoregressive conditional heteroskedasticity specifications known as TARCH and EGARCH. We also test for asymmetry after controlling for the effect of macroeconomic factors on stock market returns using TAR and M-TAR specifications within a VAR framework. Our results show that the conditional variance is an asymmetric function of past innovations raising proportionately more during market declines, a phenomenon known as the leverage effect. However, when we control for the effect of changes in macroeconomic variables, we find no significant evidence of asymmetric behaviour of the stock market returns. There are some signs that the Portuguese Stock Market tends to show somewhat less market efficiency than other markets since the effect of the shocks appear to take a longer time to dissipate.

\end{abstract}

\begin{keyword}
Asymmetric conditional volatility, stock market returns, threshold adjustment, vector autoregression
\end{keyword}

\end{frontmatter}

\section*{Introduction}

It is well known in the financial literature that stock market index returns
are positively autocorrelated, especially when dealing with high frequency
data. This fact has been often attributed to non-synchronous trading
phenomena (see e.g. Fischer \cite{Fisher1966}, Scholes and Williams \cite%
{ScholesWilliams1977}, and Lo and Mackinlay \cite{LoMackinlay1990}), and to
time-varying short-term expected returns, or risk premia in the sense of
Fama and French \cite{FamaFrench1988}, and Conrad and Kaul \cite{Conrad1988}%
. It is also acknowledged in the empirical literature that the unconditional
distribution of returns appears to be excessively leptokurtic for many stock
market index prices. Hamao \textit{et al.}\cite{HamaoMasulisNg1990}, and Wei 
\textit{et al.}\cite{WeiChaung1995}, for example, found the presence of
excess kurtosis in the normalized residuals in virtually all mature and
emerging stock markets. Mandelbrot \cite{Mandelbrot1963}, and Fama \cite%
{Fama1965} define excess kurtosis when the return series exhibit the
fat-tailed feature and leptokurtosis when there is observable greater excess
kurtosis relatively to the normal distribution. Leptokurtosis may pose a
number of problems when trying to analyse empirically the dynamic behaviour
of stock markets. In order to deal with these problems, several authors have
attempted to use more general formulations of the distributions used in
empirical work (see for example Mandelbrot \cite{Mandelbrot1963}, Fama \cite%
{Fama1965}, and Nelson \cite{Nelson1991}). However, the subject appears to
be far from solved in a totally satisfying manner.

There are several papers that are studying these topics by applying
different ideas and techniques (\cite{Boston}; \cite{Masoliver}; \cite%
{Bouchaud2001}; \cite{EnglePatton}; \cite{Bouchaud2002}; \cite{Muzy}; \cite%
{Mandelbrot etal.}; \cite{Bacry}).

Another issue obviously related to the previous one concerns the fact that
short-term stock returns usually exhibit volatility clustering. Links
between information and volatility of financial assets are of course also
central to financial economics. Ross \cite{Ross1989}, for example, argued
that volatility could be regarded as a measure of the information flow.
Moreover, as noted above, empirical researchers also report significant
relationships between flows and volatility.

Volatility clustering and excess kurtosis may arise in the context of
asymmetric behaviour of price index movements or their corresponding
returns. Such asymmetry may depend on other economic and financial factors
or simply be the result of the time-path behaviour of the price index series
itself. In the latter case, the conditional variance may be an asymmetric
function of past innovations raising proportionately more during market
declines. This means that changes in stock prices or returns tend to be
negatively related to changes in volatility (see e.g. Black \cite{Black1976}%
, and Christie \cite{Christie1982}), \textit{i.e.} the so-called leverage
effect, where stock price declines increase the financial leverage and,
consequently, the degree of risk (volatility).

Authors such as Engle \cite{Engle1993}, \textit{inter alia}, defend the idea
that financial market volatility can be predictable. In order to forecast
the volatility, several methods have been proposed in the specialised
literature, many of which are based on regression-like models. Clearly, the
classical regression model is unsuited when the residuals exhibit features
that depart from the basic OLS assumptions. Non-normality, autocorrelation,
and heteroskedasticity are some of the problems that are typically present
in this type of data. In order to deal with the problems of autocorrelation
and heteroskedasticity, ARCH/GARCH specifications are usually well suited.
However, they assume that the conditional variance has a symmetric behaviour
and thus may not fully capture the issues of non-normality. In our context,
we shall use TARCH and EGARCH model specifications to assess the extent of
non-linear dynamic behaviour underlying the stock prices data of various
markets. The models used in this paper belong to special classes of
non-linear models that usually generalise the more traditional
autoregressive conditional heteroskedasticity models.

Autoregressive conditional heteroskedasticity (ARCH) models have been used
quite frequently in applied financial research work. Among the most popular
extensions are the Threshold ARCH (TARCH) model proposed by Zakoian \cite%
{Zakoian1990}, and Glosten \textit{et al.} \cite{Glosten1993}, and the
Exponential GARCH (EGARCH) model proposed by Nelson \cite{Nelson1991}. Engle
and Ng \cite{EngleNg1993} indicate that these latter volatility models
usually outperform standard Generalized ARCH (GARCH) models.

An alternative way for modelling the volatility of stock market returns is
based on a Vector Autoregressive (VAR) specification using other variables
that may help to explain the behaviour of financial markets. In this context
we could observe e.g. (Sims [\cite{Sims72}, \cite{Sims80a} and \cite{Sims80b}%
]).

A VAR system is a reduced form model that accounts for the effect of lagged
values of the endogenous variables on the dynamic behaviour of the whole
system. Each variable is specified as a function of the past values of the
other (endogenous) variables but does not deppend on their contemporary
values. VAR models are well suited to deal with situations where an error
correction mechanism (ECM) can describe the behaviour of the system. In this
case a corresponding Vector Error Correction model (VECM) can be specified
and fully estimated using ML procedures. This is important when one needs to
separate the long-run and short-run components of the dynamic mechanism.

However, there are several criticisms with respect to VAR models. One is the
likely loss in efficiency due to lack of \textit{apriori} information.
Another problem relates with the large number of parameters to be estimated
as the dynamic structure of the system increases. This is especially \
relevant for small data sets. Other problems may arise when dealing with
impulse response analysys.

Since stock market returns and other macroeconomic variables may all be
endogenous in the system, the VAR specification is a quite satisfactory way
for modelling the volatility. Testing for asymmetries in this context
entails the use of TAR and M-TAR tests of the VAR residuals (see e.g. Enders
and Granger \cite{EndersGranger1998}, and Enders and Siklos \cite%
{EndersSiklos2001}). It is important, therefore, to obtain comparative
results of the relative performance of these two alternative types of
volatility models, and this is the main aim of the present paper.

\section{Asymmetry and Volatility Models}

The trade-off between risk and expected return is a foundation of modern
financial theory, as can be seen in the Capital Asset Pricing Model (CAPM),
Arbitrage Pricing Model (APT), Portfolio Theory, and Option Pricing Theory.
In all these models, volatility plays a central role in determining the fair
value of a security. To quantify the risk level, several researchers have
used quantitative statistical indicators, where the standard deviation is
one of the most popular measures. This statistical measure, however, assumes
a symmetric variation of price levels. Yet, asymmetries are easily observed
in time series of financial data, and already constitute an important
phenomenon.

For example, there are good reasons to believe that speculative financial
asset changes have in general an asymmetric behaviour. As mentioned above,
the leverage effect was found in many empirical studies that analyse the
behaviour of stock returns. This circumstance highlights the need for using
asymmetric models when one is analysing data on stock market behaviour.
Asymmetric behaviour in financial data can be detected with relatively ease,
since volatility raises more for negative shocks than for positive shock
with the same amplitude. In order to account for this phenomenon, two
extensions of the basic GARCH model can be used, among other possibilities:
the Threshold Autoregressive Conditional Heteroskedasticity (TARCH) and the
Exponential Generalised Autoregressive Conditional Heteroskedasticity
(EGARCH) models (see Nelson \cite{Nelson1991}, Zakoian \cite{Zakoian1990},
and Glosten \textit{et al.} \cite{Glosten1993}).

Additionally, this study also investigates the relationship between the
volatility terms for index returns and two macroeconomic variables: Dividend
Yield (Dy) and the Price Earnings per Share (PER). These variables will be
used for TAR and M-TAR tests based on a VAR model specification as explained
above. This approach is similar to that used by Gjerde and Saettem \cite%
{GjerdeSaettem1999} and Rapach \cite{Rapach2001}. Note that the daily
frequency of the stock market returns variable imposes restrictions on the
choice and availability of other macroeconomic variables that could also be
of relevance but are not available with such a frequency.

\subsection{The Data}

The data used in this study consists of daily stock index closing prices for
seven stock market indices over a period from January 1993 until December
2002 (Datastream database). Stock returns were computed as the logarithm of
the ratio of the index values at time $t$ and time $t-1$. We consider the
following indices: the Standard and Poors 500 (S\&P500) to represent the
whole US economy, the CAC 40 index for France, the FTSE 100 for the UK, the
DAX for Germany, the IBEX 35 for Spain, the ASE 20 for Greece and finally,
the PSI 20 for Portugal.

The exchanges of Paris, London, Frankfurt, and New York represent four of
the world's major centres for the trading and distribution of both domestic
and international equities with special interest for Portuguese investors,
and this justifies their inclusion in the analysis. Furthermore, France and
Germany are viewed as \textquotedblleft core\textquotedblright\ European
economies, being members of the Exchange Rate Mechanism since its inception
in 1979, and of the European common currency - the Euro - since 1999. The
remaining stock markets are more similar to the Portuguese stock market and
we may expect somewhat similar results for the three southern European
countries.

\subsection{Methodological Issues and Empirical Results}

The existence of important inter-relationships between the major financial
stock markets gives support to the use of VAR systems in the context of the
analysis of volatility. There is the need to understand how shocks and
volatility in one market are transmitted to other markets. To this end, we
need to look at the extent to which multi-lateral interaction exists between
these markets. Thus, after identifying the channels of interaction we can
observe the implications of volatility co-movements between different
markets.

VAR models have been increasingly used in the analysis of the relationships
between financial markets. VAR models may be used with both stationary and
non-stationary data. In the latter case, they provide a convenient framework
to test for cointegration using the Johansen methodology, but the
methodology is far more general and is also quite useful when the variables
are stationary. On the other hand, ARCH/GARCH type models have been widely
used to test hypotheses concerning the conditional volatility of stock
market returns. Both methods can be employed to model volatility in the
stock market. However, it is important to note that employing these
different models, each one measures and captures different types of
volatility.

TARCH and EGARCH models are especially adequate to model the volatility
measured in terms of error's variance in the context of asymmetry. On the
other hand, TAR and M-TAR models systematise the volatility of the dependent
variables in a VAR system, captured from the corresponding residuals. The
TAR and M-TAR models based on VAR systems are especially designed to capture
asymmetric multivariate effects in a multi-equation framework, which
certainly constitutes an important advantage over the single-equation
multivariate TARCH and EGARCH models. However, one important disadvantage of
the former models is that they do not account for conditional
heteroskedasticity that may be present in the data.

The first step of our analysis consists of estimating univariate TARCH(1,1)
and EGARCH(1,1) models in order to assess whether our series exhibit or not
some type of conditional volatility asymmetric behaviour. For the TARCH
model we use the following specification of the conditional variance:%
\begin{equation}
\sigma _{t}^{2}=\omega +\alpha \varepsilon _{t-1}^{2}+\gamma \varepsilon
_{t-1}^{2}d_{t-1}+\beta \sigma _{t-1}^{2}  \label{sigma}
\end{equation}%
where the variance $\sigma _{t}^{2}$ is a function of the past squared
residuals $\varepsilon _{t-1}^{2}$, and of its own lagged values $\sigma
_{t-1}^{2}$. The variable $d_{t-1}$ is a dummy variable equal to one if $%
\varepsilon _{t-1}>0$, and equal to zero otherwise. $\omega $, $\alpha $, $%
\gamma $, and $\beta $ are the parameters of the conditional variance
equation that will be estimated. The equation of the conditional variance
for the EGARCH model takes the following form: 
\begin{equation}
\log \sigma _{t}^{2}=\omega +\beta \log \sigma _{t-1}^{2}+\alpha \left\vert 
\frac{\varepsilon _{t-1}}{\sigma _{t-1}}\right\vert +\gamma \frac{%
\varepsilon _{t-1}}{\sigma _{t-1}}
\end{equation}%
where the symbols are as described earlier. The results of the univariate
TARCH and EGARCH models are reported in Table 1.

As can be seen, the estimates of $\alpha $ are almost all significantly
positive, except for the US and UK, where we found no significant $\alpha $
coefficients when the TARCH model was used. In this model, we obtain good
news $(\varepsilon _{t}\leq 0)$ and bad news $(\varepsilon _{t}>0)$. This
means that the model has differential effects on the conditional variance --
good news has an impact on $\alpha $, while bad news has an impact on $%
\alpha +\gamma $. We say that the leverage effect exists if $\gamma \neq 0$
and an asymmetric effect is observed if $\gamma >0$. The leverage effect
expressed in $\varepsilon _{t-1}^{2}d_{t-1}$ is associated with the
parameter $\gamma $ in equation (\ref{sigma}), and at the same time it must
reveal statistical significance for both models in Table 1. In the EGARCH
specification, the asymmetric behaviour exists if $\gamma <0$.

Another important result to note is that the estimate of $\beta $ in the
EGARCH model is less than one for all the markets, which implies that all
the moments of the relevant statistical distribution exist and that the
quasi-maximum likelihood estimators are likely to be consistent and
asymptotically normal. Thus, relatively to the results reported in Table 1,
we may conclude that the leverage effect hypothesis cannot be rejected in
any case. Likewise, there is evidence of an asymmetric behaviour of the
stock markets for all the countries analysed.

\begin{center}
\bigskip $\FRAME{itbpFU}{4.1745in}{6.052in}{0in}{\Qcb{Table 1. TARCH(1,1)
and EGARCH(1,1) }}{}{table 1.wmf}{\special{language "Scientific Word";type
"GRAPHIC";maintain-aspect-ratio TRUE;display "USEDEF";valid_file "F";width
4.1745in;height 6.052in;depth 0in;original-width 7.5403in;original-height
10.9537in;cropleft "0";croptop "1";cropright "1";cropbottom "0";filename
'Table 2.wmf';file-properties "XNPEU";}}$
\end{center}

A more detailed analysis of the results reported in Table 1 reveals that the
coefficient $\alpha $ for Portugal is higher than the corresponding values
for the remaining markets, both for the TARCH and EGARCH models\footnote{%
Except for Greece in the EGARCH model, whose value of $\alpha $ is $0.2602,$
whereas for Portugal is $0.2349.$} \ These figures imply that shocks in the
Portuguese stock market tend to have longer durations, or higher
persistence. As Bala and Prematne \cite{Bala2003} state, this circumstance
implies that the effect of shocks in earlier periods for Portugal tend to
linger around for a longer period than it does in other stock markets. One
possible explanation for this result is that it may imply that the
Portuguese stock market shows less market efficiency than the other markets
analysed, since the effects of shocks take longer time to dissipate.

However, one should note that this does not mean that the Portuguese stock
market is not efficient. Our results should just be analysed in relative
terms and comparatively with other markets.

We turn now to consider the results of multivariate TARCH and EGARCH models
where the Portuguese stock market index is the dependent variable, in order
to obtain the OLS residuals, and take the remaining market indices
successively as the independent market index for all possible relations
established with the Portuguese stock market. These results may bring new
information relatively to the results of Table 1, since it is quite likely
that the stock markets establish interactions between them. We restrict our
analysis to the interactions between the Portuguese stock market and the
remaining markets, successively, and we estimate multivariate TARCH and
EGARCH models. The results reported in Table. 2 are very similar to those
reported earlier and lead in general to the same conclusions. The estimated
coefficients, however, appear now significant in all cases.

We are now going to test the possibility that an adjustment process occurs
for different stock market price index combinations with the Portuguese
index, which have asymmetric cycles in the respective time series. For this
test we employ the methodology of TAR and M-TAR models, where the indices
appear as endogenous variables and the macroeconomic factors as control
exogenous variables.

The basic starting point for this analysis is the long-run relationship
between the relevant variables to the model. The long-run equilibrium
relationship between two time series $z_{t}$ and $x_{t}$ can be estimated as
a standard regression model $z_{t}=\alpha +\beta x_{t}+\mu _{t}$, where $%
\alpha $ and $\beta $ are the estimated parameters, and $\mu _{t}$ is a
disturbance term that may be serially correlated. The parameter $\beta $
gives the magnitude of adjustment of $z$ to variations in $x$, and is the
long-run elasticity of the two variables if they are measured in logs. If $%
\beta <1$, then shifts in $x$ are not fully passed onto $z$.

The second step of the methodology focuses on the OLS estimates of $\rho
_{1} $ and $\rho _{2}$ in the following error correction model:%
\begin{equation}
\Delta \mu _{t}=I_{t}\rho _{1}\mu _{t-1}+(1-I_{t})\rho _{2}\mu
_{t-1}+\varepsilon _{t}  \label{ecm}
\end{equation}%
where $\varepsilon _{t}$ is a white noise disturbance and the residuals from
the long-run equation are used to estimate $\Delta \mu _{t}$. $I_{t}$ is the
Heaviside indicator function such that%
\begin{equation}
I_{t}=\left\{ 
\begin{array}{c}
1\text{ \ if \ }\xi _{t-1}\geq \tau \\ 
0\text{ \ if \ }\xi _{t-1}<\tau%
\end{array}%
\right. .  \label{heaviside}
\end{equation}

\begin{center}
\begin{equation*}
\FRAME{itbpFU}{4.5567in}{6.6072in}{0in}{\Qcb{Table 2. TARCH and EGARCH
multivariate results}}{}{table 2.wmf}{\special{language "Scientific
Word";type "GRAPHIC";maintain-aspect-ratio TRUE;display "USEDEF";valid_file
"F";width 4.5567in;height 6.6072in;depth 0in;original-width
7.5403in;original-height 10.9537in;cropleft "0";croptop "1";cropright
"1";cropbottom "0";filename 'Table 2.wmf';file-properties "XNPEU";}}
\end{equation*}
\end{center}

If in $(4)$ $\xi _{t-1}=\mu _{t-1}$, then the model specification
illustrated in (\ref{ecm}) is called the threshold autoregressive (TAR)
model. It allows for different coefficients of positive and negative
variations. A sufficient condition for the stationarity of $\mu _{t}$ is $-2$
$<$ ($\rho _{1}$,$\rho _{2}$)$<0$. This means that the long-run equation is
an attractor such that $\mu _{t}$ can be written as an error correction
model similar to that given in (\ref{ecm}). If $\rho _{1}=\rho _{2}$ then
the adjustment is symmetric, which is a special case of (\ref{ecm}) and (\ref%
{heaviside}). Expression (\ref{ecm}) can also contain lagged values of $%
\Delta \mu $. When $\mu _{t}$ is above its long-run equilibrium value, the
adjustment is $\rho _{1}\mu _{t-1}$, and if $\mu _{t-1}$ is below its
long-run equilibrium value, the adjustment is $\rho _{2}\mu _{t-1}$.

If in (\ref{heaviside}) $\xi _{t-1}=\Delta \mu _{t-1}$, then the model (\ref%
{ecm}) is called the momentum threshold autoregressive (M-TAR) model. The
M-TAR model allows the decay to depend on the previous period change in $\mu
_{t-1}$. The value of the threshold $\tau $, in our case, will be assumed to
be zero in all models.

The TAR model is designed to capture asymmetrically deep movements in the
series of the deviations from the long-run equilibrium, while the M-TAR
model is useful to capture the possibility of asymmetrically steep movements
in the series (Enders and Granger \cite{EndersGranger1998}). For example, in
the TAR \ model if $-1<$ $\rho _{1}<\rho _{2}$ $<0$, then the negative phase
of $\mu _{t}$ will tend to be more persistent than the positive phase. On
the other hand, for the M-TAR model, if for example \TEXTsymbol{\vert}$\rho
_{1}|<|\rho _{2}|$ the model exhibits little decay for positive $\Delta \mu
_{t-1}$ but substantial decay for negative $\Delta \mu _{t-1}$. This means
that increases tend to persist but decreases tend to revert quickly toward
the attractor.

Finally, we can perform a number of statistical tests on the estimated
coefficients (and also on the residuals) in order to ascertain the validity
of the error correction model outlined in (\ref{ecm}), and subsequently if
the adjustment is symmetric or not. The relevant tests on the coefficients
are $H_{0}:\rho _{1}$ $=0$ and $H_{0}:\rho _{2}$ $=0$, for which we obtain
the sample values of the t-statistics; and $H_{0}:\rho _{1}$ $=\rho _{2}=0$
, for which we obtain the sample values of the $F$-statistic. The
restriction that adjustment is symmetric, i.e. $H_{0}:\rho _{1}=\rho _{2}$,
can also be tested using the usual $F$-statistic.

If the variables in the long-run equation are stationary, the usual critical
values of the $t$ and $F$ distributions can be used to assess the
significance level of the underlying tests. However, if these variables are
integrated of first order, one can use the critical values reported by
Enders and Siklos \cite{EndersSiklos2001} to determine whether the null
hypothesis of no cointegration can be rejected. If the alternative
hypothesis is accepted, it is possible to test for asymmetric adjustment
using the standard critical values of the $F$ distribution, since $\rho _{1}$
and $\rho _{2}$ converge to a multivariate normal distribution (Enders and
Granger \cite{EndersGranger1998}).

Table 3 contains the results of the TAR and M-TAR models. As may be seen,
for the first two tests the null hypothesis is rejected in all cases. Thus,
we may proceed our analysis with the tests of symmetry and the results are
that the null hypothesis is not rejected in any of the models considered.

\begin{center}
$\FRAME{itbpFU}{5.1206in}{6.3114in}{0in}{\Qcb{Table 3. TAR and M-TAR results}%
}{}{table 3.wmf}{\special{language "Scientific Word";type
"GRAPHIC";maintain-aspect-ratio TRUE;display "USEDEF";valid_file "F";width
5.1206in;height 6.3114in;depth 0in;original-width 7.5723in;original-height
9.3417in;cropleft "0";croptop "1";cropright "1";cropbottom "0";filename
'Table 3.wmf';file-properties "XNPEU";}}$
\end{center}

\section{ Conclusions}

This paper presents a comparative analysis between four volatility models
(TARCH, EGARCH, TAR and M-TAR) in order to ascertain their power to capture
asymmetric cycles in the underlying time series. Our results indicate that
while the TAR and M-TAR models do not identify the presence of asymmetry in
any market, the TARCH and EGARCH models have captured asymmetry in all cases
under analysis.

Another relevant conclusion from the multivariate TARCH and EGARCH results
is that the Portuguese stock market appears to reveal higher $\alpha $
values than the remaining markets, for both models. These figures imply that
shocks in the Portuguese market, in the later periods, tend to have longer
duration periods. This circumstance implies that the effects of shocks in
earlier periods tend to linger around for a longer period than it does in
other stock markets. This means that for the Portuguese stock market the
effects of shocks take a longer time to dissipate.

\end{document}